\documentclass[11pt]{article}
\usepackage{epsfig}
\newcommand{\beq}{\begin{equation}}
\newcommand{\eeq}{\end{equation}}
\newcommand{\bea}{\begin{eqnarray}}
\newcommand{\eea}{\end{eqnarray}}
\newcommand{\lsim}{\mathrel{\mathop{\kern 0pt \rlap
  {\raise.2ex\hbox{$<$}}}
  \lower.9ex\hbox{\kern-.190em $\sim$}}}
\newcommand{\gsim}{\mathrel{\mathop{\kern 0pt \rlap
  {\raise.2ex\hbox{$>$}}}
  \lower.9ex\hbox{\kern-.190em $\sim$}}}

\begin{document}
\title{\bf Weak non-mesonic decay of Hypernuclei}
\author{W.M. Alberico\footnote{Presented at XXVIII Mazurian Lakes Conference,
Krzyze, Poland, August 31-September 7, 2003}
\vspace*{0.3cm}\\
\begin{tabular}{c}
{\it Dipartimento di Fisica Teorica and INFN,}\\ 
{\it via P. Giuria 1, I-10125 Torino, Italy}
\end{tabular}
}
\date{\today}
\maketitle
\begin{abstract}
We review the mechanism of weak decay of hypernuclei, with emphasis on
the non-mesonic decay channels. Various theoretical approaches are discussed
and the results are compared with the available experimental data.
\end{abstract}
\vspace{0.8cm}
  
\section{Introduction}
The main decay modes of $\Lambda$-hypernuclei are the so-called mesonic 
and non-mesonic decays(for a recent review on the argument see 
ref.~\cite{AlbGar}): the former also occurs for free $\Lambda$-hyperons:
\beq
\label{lambdadec}
\Lambda \rightarrow \left\{
\begin{array}{l l }
 \pi^- p \qquad\qquad & (\Gamma^{\rm free}_{\pi^-}/
\Gamma^{\rm free}_{\Lambda}=0.639)  \\
\pi^0 n  & (\Gamma^{\rm free}_{\pi^0}/
\Gamma^{\rm free}_{\Lambda}=0.358) , 
\end{array}\right.
\eeq
 The experimental ratio of the relevant widths, 
$\Gamma^{\rm free}_{\pi^-}/\Gamma^{\rm free}_{\pi^0}\simeq 1.78$,
together with the measurements of $\Lambda$ polarization observables
lead to the formulation of the $\Delta I=1/2$ rule on the isospin change of
the system [from simple Clebsch-Gordon coefficient analysis, the latter 
would predict $\Gamma^{\rm free}_{\pi^-}/\Gamma^{\rm free}_{\pi^0}=2$]. 
This rule is based on experimental observations, but its dynamical origin 
is not yet understood on theoretical grounds. 
The Standard Model does not support it and many non-perturbative effects could
be responsible for the measured enhancement of the $\Delta I=1/2$ transition 
amplitude.

The mesonic decay can also take place in  $\Lambda$-hypernuclei, however the
Pauli principle tends to disfavour the produced final state, with an emitted
nucleon of momentum $p_N\simeq 100$ MeV/c (from a $\Lambda$ at rest), 
well below
the maximum level of occupied states in the nucleus. Of course this argument 
is only a qualitative one: indeed the mesonic decay is an important fraction 
of the total decay rate in light-medium  $\Lambda$-hypernuclei. This
outcome stems from several facts: {\it i)} the hyperon momentum distribution 
in the nucleus, {\it ii)} the medium attraction on the emitted pion, which 
lowers its mass at fixed momentum and distorts the pionic outgoing wave,
{\it iii)} the strong reduction of the local Fermi momentum at the nuclear
surface.\cite{Nieves}

Notably the mesonic decay can be used to extract information on the 
pion--nucleus optical potential (in a complementary way with respect to
pionic atoms and low energy $\pi$-nucleus scattering experiments).

The  $\Lambda$-hypernuclei non-mesonic decay occurs via the weak 
interaction of the $\Lambda$ with one nucleon:
\[
\Lambda N \to NN \qquad\qquad \mbox{ {(one-body induced decay)}} \]
or with a pair of correlated nucleons:
\[
\Lambda NN \to NNN \qquad\qquad \mbox{ {(two-body induced decay)}}. \]

\begin{figure}[ht]
\begin{center}
\mbox{\epsfig{file=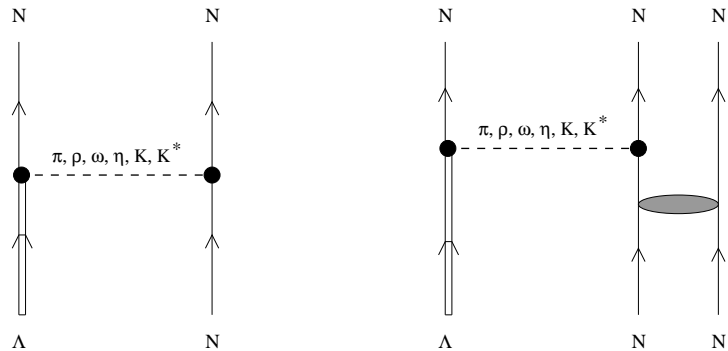,width=0.75\textwidth}}
\end{center}
\caption{
One--nucleon and two--nucleon induced $\Lambda$ decay in nuclei}
\label{nm12}
\end{figure}

These processes are mediated by the exchange of a meson (including the 
$K$, $K^*$ strange mesons), as illustrated in Fig.~\ref{nm12}

%

The decay width associated to the various non-mesonic processes can be denoted
as follows ($n$ indicates a neutron, $p$ a proton):
\begin{eqnarray}
\label{gn}
&\Gamma_n: \hspace{4mm}   \Lambda n & \rightarrow  nn \\
\label{gp}
&\Gamma_p: \hspace{4mm} \Lambda p & \rightarrow  np \\
\label{g2}
&\Gamma_2: \hspace{3mm} \Lambda NN & \rightarrow  NNN 
\end{eqnarray}
whence the total non-mesonic width is 
$\Gamma_{\rm NM}=\Gamma_1+\Gamma_2\equiv \Gamma_n+\Gamma_p+\Gamma_2$.
The total hypernuclear decay width is then:
\beq
\Gamma_{\rm T}=\Gamma_{\rm M}+\Gamma_{\rm NM}
\label{gammatot}
\eeq
where $\Gamma_{\rm M}=\Gamma_{\pi^-}+\Gamma_{\pi^0}$ is the mesonic width.

The non-mesonic mode can occur {\it only} in nuclei, and therefore it is 
the only source of information on the ${\Lambda}N \rightarrow NN$ weak 
interaction.

Concerning the kinematics of the process, it is worth noticing that if one 
assumes the $\Lambda$ at rest ($Q$-value of about 
176~MeV$\simeq m_{\Lambda}-m_N$) and an equal share of the available energy 
between the two (respectively, three) final nucleons, the momenta of the 
emitted nucleons are expected of the order of $p_N\simeq 420$~MeV/c for the
one-body induced decay (respectively, $p_N\simeq 340$~MeV/c for the two-body
induced decay). Hence we do not expect a significative influence of the
Pauli principle on these decay channels.

An interesting feature of the experimental decay widths of 
$\Lambda$-hypernuclei is the approximate stability of this quantity in 
passing from medium-light to very heavy systems: the mesonic ($\Gamma_{\rm M}$)
and non-mesonic ($\Gamma_{\rm NM}$) partial widths depend upon the mass number 
in such a way as to compensate each other. This saturation property is clearly
related to the short range of the  ${\Lambda}N $ weak 
interaction.\cite{Cohen,OsRa}

\section{Theoretical models for $\Lambda$-hypernuclear decay}

The starting point for all methods described below is the 
weak effective hamiltonian for ${\Lambda}\to \pi N$ decay: 
\beq
{\mathcal H}^W_{{\Lambda}\pi N}=iG m_{\pi}^2\overline{\psi}_N(A+B\gamma_5)
{\vec \tau} \cdot {\vec \phi}_{\pi}{\psi}_{\Lambda} ,
\label{Hweak}
\eeq
the parameters of which are fixed on the free  $\Lambda$ decay:
$G= 2.211\cdot 10^{-7}/m_{\pi}^2$, $A=1.06$ (PV amplitude) and
$B=-7.10$ (PC amplitude). To enforce the  $\Delta I=1/2$ rule 
the hyperon is assumed to be an isospin spurion with $I=1/2$, $I_z=-1/2$.

In the non-relativistic limit, one can then express the {\it free} 
$\Lambda$ decay width as follows:
\bea
&&\Gamma^{\rm free}_{\alpha}=c_{\alpha}(G m^2_{\pi})^2\int
\frac{d\vec q}{(2\pi)^3\,2\omega({\vec q})}\,2\pi\,
\nonumber\\
&&\,\quad\qquad\times\delta[E_{\Lambda}-\omega({\vec q})-E_N]
 \left(S^2+\frac{P^2}{m^2_{\pi}}{\vec q}^2\right) ,
\label{Gammafree}
\eea
where $c_{\alpha}=1(2)$ for $\Gamma_{\pi^0}$ 
($\Gamma_{\pi^-}$, respectively) and the decay occurs both through 
parity violating (the  s-wave amplitude, $S=A$) and parity conserving 
(the  p-wave amplitude, $P=m_{\pi}B/(2m_N)$) terms.

\subsection{The wave function method}

An expression similar to (\ref{Gammafree}) for the {\bf mesonic width} of 
hypernuclei can be obtained by explicitly taking into account the 
hyperon ($\phi_{\Lambda}$), pion ($\phi_{\pi}$) and nucleon ($\phi_N$) 
wave functions inside the nucleus~\cite{Nieves,Motoba,Parreno}:

\begin{eqnarray}
&&\!\!\!\!\!\!
 \Gamma_{\alpha}=c_{\alpha}(G m^2_{\pi})^2\sum_{N\ne F}\int
\frac{d\vec q}{(2\pi)^3\,2\omega({\vec q})}\,2\pi\,
\delta[E_{\Lambda}-\omega({\vec q})-E_N]\times 
\label{Gammawf} \\
&&\times \left\{S^2\left|\int d{\vec r} \phi_{\Lambda}(\vec r)
\phi_{\pi}({\vec q}, {\vec r})\phi^*_N(\vec r)\right|^2
+\frac{P^2}{m^2_{\pi}}
\left|\int d{\vec r} \phi_{\Lambda}(\vec r){\vec \nabla}
\phi_{\pi}({\vec q}, {\vec r}) \phi^*_N(\vec r)\right|^2 \right\} 
\nonumber
\end{eqnarray}
The pion wave function corresponds to an outgoing wave solution of 
the Klein--Gordon equation in the presence of a suitable $\pi$-nucleus 
optical potential, $V_{\rm opt}$, while the  $\Lambda$ wave function can
be derived within a variety of hypernuclear shell models, whose parameters
are typically determined by comparison with the available spectroscopic 
data.

Turning now to the evaluation of the {\bf non-mesonic} width, one needs an 
explitic model for the $\Lambda N\to NN$ weak transition: the latter is 
usually described in terms of the exchange of virtual mesons belonging to
the pseudoscalar ($\pi,\eta$ and $K$) or to the vector ($\rho, \omega$
and $K^*$) octets. The most important (from an heuristic point of view) 
component of the $\Lambda N\to NN$ transition potential is associated to 
the exchange of a pion: 
\beq
V_{\pi}({\vec q})=-G m_{\pi}^2\frac{g_{NN\pi}}{2m_N}\left(
A+\frac{B}{2\bar{m}}{\vec \sigma_1} \cdot {\vec q}\right)
\frac{{\vec \sigma_2} \cdot {\vec q}}{{\vec q}^2+m_{\pi}^2} 
{\vec \tau_1} \cdot {\vec \tau_2}
\label{Vpion}
\eeq
with $\bar{m}=(m_{\Lambda}+m_N)/2$.

This is the only component of the  $\Lambda N\to NN$ meson exchange potential,
for which the couplings of both the weak and strong vertices are well 
constrained by the existing phenomenology and experimental data. The exchange
of heavier mesons, which cannot be produced in the free-$\Lambda$ weak decay,
is subject to theoretical uncertainties and to a somewhat large 
indetermination of the model parameters (typically couplings and masses), in 
spite of the existence of various theoretical schemes which, in principle,
should allow to fix them on a firm basis. Nevertheless the influence of 
$\Lambda N\to NN$ processes mediated by heavier mesons on the calculated 
non-mesonic decay widths appears to be important in order to reproduce the 
available experimental data.

A sort of ``minimal model'', which has been often employed in the 
literature~\cite{RaOsSa}, accounts for the exchange of pions and $\rho$-mesons,
together with phenomenological, $q$-dependent short range correlations, 
usually parameterized in the Landau-Migdal form (see, for details, 
ref.~\cite{AlbGar}).

In the framework of the wavefunction method, the one-body induced 
non-mesonic decay width takes the form:
\begin{equation}
\label{nm-wfm}
\Gamma_1=\int \frac{d{\vec p}_1}{(2\pi)^3}\int \frac{d{\vec p}_2}{(2\pi)^3}
\,2\pi\, \delta({\rm E.C.}) \overline{{\sum}}
\left|{\mathcal M}({\vec p}_1,{\vec p}_2)\right|^2 ,
\end{equation}
where
\beq
{\mathcal M}({\vec p}_1,{\vec p}_2)
\equiv \langle \Psi_R; N({\vec p}_1)N({\vec p}_2)|
\hat{T}_{\Lambda N\to NN}|\Psi_H\rangle
\eeq
is the transition matrix element between the initial hypernuclear state 
$|\Psi_H\rangle$ and the final state with two outgoing nucleons, while
$\delta({\rm E.C.})$ guarantees energy conservation.
Besides the above mentioned uncertainties in the OME (One-Meson-Exchange) 
transition potential, one faces here also the complications associated with
a strongly interacting many-body system.

We will not consider here further details of this method, however it is worth
recalling that the wave function approach is considered to be the most 
appropriate one for the calculation of light $\Lambda$-hypernuclei mesonic 
widths; it is, instead, less accurate for the non-mesonic widths and 
certainly less practical for the heavier systems.

\subsection{The polarization propagator method}

This method obtains the hypernuclear decay width through the evaluation 
of the $\Lambda$ self-energy inside the nuclear 
medium~\cite{OsSa,ADEM,RaOsSa,ADGR,ADGC}:
\begin{equation}
{\Gamma}_{\Lambda}=-2\;{\rm Im}\,{\Sigma}_{\Lambda}\,.
\label{self0}
\end{equation}

\begin{figure}
\begin{center}
\mbox{\epsfig{file=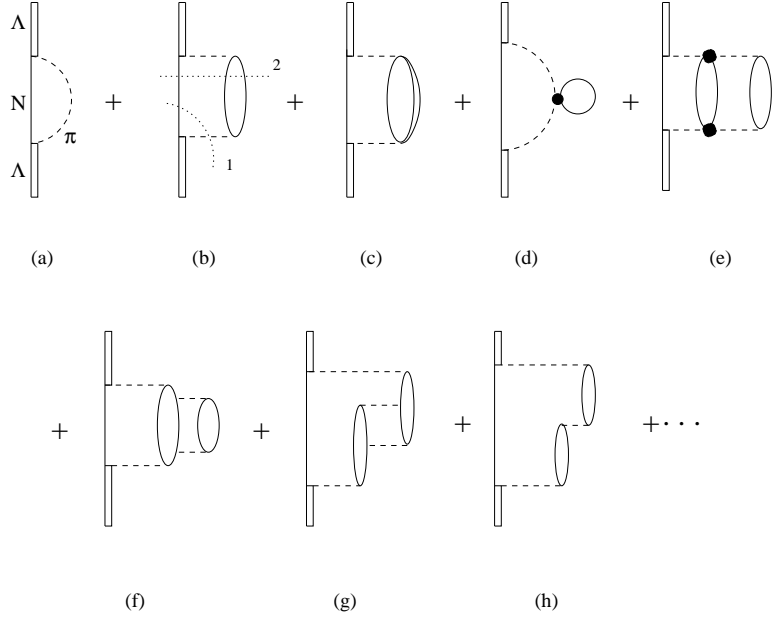,width=0.8\textwidth}}
\vskip 2mm
\caption{
Lowest order terms for the ${\Lambda}$ self--energy in nuclear matter.}
\label{self1}
\end{center}
\end{figure}

The starting point is again the weak effective Hamiltonian (\ref{Hweak}), 
which provides the relevant vertices already in the simplest process 
illustrated in diagram (a) of Fig.~\ref{self1}, which corresponds to the 
free $\Lambda$ decay. More generally, a few  contributions to the
$\Lambda$ self-energy in the medium are illustrated in diagrams (b)--(h)
 of Fig.~\ref{self1}, where the pion propagator is typically dressed by 
the strong interaction with particle-hole (ph), $\Delta$-hole ($\Delta$h)
and two particles-two holes (2p2h) intermediate states. Further contributions
arise from the medium modifications on the nucleon propagator, 
which has origin in the $\Lambda\pi N$ weak vertex.

Formally the $\Lambda$ self-energy can be written as:
\beq
{\Sigma}_{\Lambda}(k)=3i(G m_{\pi}^2)^2\int \frac{d^4q}{(2\pi)^4}
\left(S^2+\frac{P^2}{m_{\pi}^2}\vec q\,^2\right)
 F_{\pi}^2(q)G_N(k-q)G_{\pi}(q)\,,
\label{self2}
\eeq
where $k$ is the $\Lambda$ momentum inside the nucleus, $G_N(p)$ and 
$G_{\pi}(q)$ are the nucleon and pion propagators (in nuclear matter):
\beq
G_N(p)=\frac{{\theta}(\mid \vec p \mid-k_F)}{p_0-E_N(\vec p)-V_N+i{\epsilon}}+
\frac{{\theta}(k_F-\mid \vec p \mid)}{p_0-E_N(\vec p)-V_N-i{\epsilon}}\,, 
\label{nuclprop}
\eeq
and
\beq
G_{\pi}(q)=\frac{1}{q_0^2-\vec q\,^2-m_{\pi}^2-{\Sigma}_{\pi}^*(q)}\,.
\label{pionprop}
\eeq
In equation (\ref{nuclprop}) the medium effects on the nucleon propagation
are phenomenologically embodied in the nuclear binding potential $V_N$, 
while in (\ref{pionprop}) the analogous effects on the pion appear in the
pion self-energy, illustrated above.

More explicitly, by inserting (\ref{self2}) into (\ref{self0}), one can 
express the $\Lambda$-hypernuclear decay width as:
\begin{eqnarray}
&&{\Gamma}_{\Lambda}(\vec k,\rho)=-6(G m_{\pi}^2)^2\int \frac{d\vec q}
{(2\pi)^3}{\theta}(\mid \vec k- \vec q \mid -k_F)
\nonumber  \\
&&\,\quad \times {\theta}(k_0-E_N(\vec k-\vec q)-V_N) 
{\rm Im}\left[{\alpha}(q)\right]_{q_0=k_0-E_N(\vec k-\vec q)-V_N}
\label{Sigma2}
\end{eqnarray}
where:
\begin{eqnarray}
&&{\alpha}(q)=\left(S^2+\frac{P^2}{m_{\pi}^2}\vec q\,^2\right)F_{\pi}^2(q)
G_{\pi}^0(q)+\frac{\tilde{S}^2(q)U_L(q)}{1-V_L(q)U_L(q)}+
 \nonumber \\
&&\,\,\quad +\frac{\tilde{P}_L^2(q)U_L(q)}{1-V_L(q)U_L(q)}+
2\frac{\tilde{P}_T^2(q)U_T(q)}{1-V_T(q)U_T(q)}
\label{Alpha}
\end{eqnarray}

In the above, to perform a realistic calculation, the effective interactions 
$\tilde{S}$, $\tilde{P}_L$, $\tilde{P}_T$,$V_L$, $V_T$ 
include $\pi$- and $\rho$-exchange plus
  short range repulsive correlations. Specifically:
 $V_L$, $V_T$ are the (strong) {\sl p-h} interaction and
include a Landau parameter $g^{\prime}(q)$, which embodies the NN short 
range repulsion. $\tilde{S}$, $\tilde{P}_L$ and $\tilde{P}_T$ correspond to 
the lines connecting the weak and strong hadronic vertices and contain 
another Landau parameter, $g^{\prime}_{\Lambda}(q)$ [a priori different from
$g^{\prime}$], which is intended to parameterize the strong $\Lambda N$ 
short range correlations.

Besides the interaction lines, the other key-ingredient of (\ref{Alpha}) 
are the longitudinal and transverse (with respect to the pion momentum 
${\vec q}$) polarization propagators, $U_L$ and $U_T$. They contain 
the Lindhard functions for {\sl p-h} and {\sl ${\Delta}$-h} excitations 
and the irreducible {\sl 2p-2h} polarization propagator:
\beq
U_{L,T}(q)=U^{ph}(q)+U^{\Delta h}(q)+U^{2p2h}_{L,T}(q)
\label{ULT}
\eeq

The imaginary part of $\alpha(q)$, needed to evaluate ${\Gamma}_{\Lambda}$,
 develops various contributions, namely:
\beq
{\rm Im}\, \frac{U_{L,T}(q)}{1-V_{L,T}(q)U_{L,T}(q)}=
\frac{{\rm Im}\,U^{ph}(q)+{\rm Im}\,U^{\Delta h}(q)+
{\rm Im}\,U^{2p2h}_{L,T}(q)}{\mid 1-V_{L,T}(q)U_{L,T}(q)\mid ^2}\,.
\label{Imalpha}
\eeq
Indeed the three terms represent different decay mechanisms of 
the hypernucleus:
\bea
&\Gamma_M \propto {\rm Im}\,U^{\Delta h}\qquad
&\mbox{(part of mesonic width)}
\nonumber\\
&\Gamma_1\, \propto {\rm Im}\,U^{ph}\qquad
&\mbox{(non-mesonic, one-body induced decay width)}
\nonumber\\
&\Gamma_2 \propto {\rm Im}\,U^{2p2h}\qquad
&\mbox{(non-mesonic, two-body induced decay width}
\nonumber\\
& &\mbox{ and additional part of mesonic width)}
\nonumber
\eea

While the {\sl p-h} and {\sl ${\Delta}$-h} polarization propagators are well
known and can be analytically evaluated~\cite{FetWal}, the {\sl 2p-2h}
polarization propagator, even in the non-relativistic limit considered here,
demands quite a computing effort to be fully determined. On the other hand,
according to the above relations, it is needed for estimating the two-body
induced decay width. 

In this context two approaches have been utilized till now 
for the evaluation of $U^{2p2h}_{L,T}(q)$:
\begin{description}
\item{A.} Phenomenological model.\\
In this case one takes into account mainly the phase space available to
2p-2h excitations in nuclear matter and determines the relevant entity of
the 2p-2h propagator through its connection with the phenomenological 
$\pi$-nucleus optical potential. The relation with specific hypernuclei is
then obtained by implementing the local density approximation.
\item{B.} Microscopic calculation.\\
This approach proceeds through the path-integral formulation and provides
the 2p-2h propagator within the so-called One Boson Loop (OBL) approximation,
which embodies a rich variety of perturbative contributions to 
$U^{2p2h}_{L,T}(q)$. 
\end{description}
\noindent
In the following we will consider only the phenomenological approach and
we refer the reader to refs.~\cite{ADGC,AlbGar} for the details and the
results of the microscopic approach.

\subsection{The phenomenological $2p2h$ propagator}

In the $(q_0,\vec q)$ region where the {\sl p-h} and {\sl $\Delta$-h} 
excitations are off--shell, the following relation between $U^{2p2h}_L$ and 
the $p$--wave pion--nucleus optical potential $V_{\rm opt}$ holds:
\beq
\frac{\displaystyle \vec q\,^2\frac{f^2_{\pi}}{m^2_{\pi}}
F^2_{\pi}(q)U^{2p2h}_L(q)}
{\displaystyle 1-\frac{f^2_{\pi}}{m^2_{\pi}}g_L(q)U_L(q)}=2q_0V_{\rm opt}(q)\,.
\label{Vopt1}
\eeq
At pion threshold the latter can be parameterized as:
\beq
2q_0V_{\rm opt}(q_0\simeq m_{\pi},\vec q\simeq \vec 0; \rho)
=-4\pi \vec q\,^2 \rho^2 C_0 
\label{Vopt2}
\eeq
where $C_0$ is a complex number  extracted from the experimental
data on pionic atoms~\cite{Garcia}; 
its ``proper part'' (using  $g'\equiv g_L(0)=0.615$)
turns out to be:
\[C^*_0=(0.105+i0.096)/m^6_{\pi}\,.\]
Hence one obtains the following parameterization of the proper {\sl 2p-2h} 
polarization propagator in the spin--longitudinal channel at pion threshold:
\beq
\vec q\,^2\frac{f^2_{\pi}}{m^2_{\pi}}F^2_{\pi}
U^{2p2h}_L(q_0\simeq m_{\pi},\vec q \simeq \vec 0;\rho)
=-4\pi \vec q\,^2 \rho^2 C^*_0\,.
\label{Vopt3}
\eeq
Further, to obtain the general dependence of $U^{2p2h}_{L,T}$ upon 
$(q_0,\vec q)$, one considers the phase space available for the real 
{\sl 2p-2h} excitations:
\begin{eqnarray}
&&P(q_0,\vec q;\rho)\propto \int \frac{d^4k}{(2\pi)^4}\,{\rm Im}\, U^{ph}
\left(\frac{q}{2}+k;\rho\right)
 \nonumber \\
&&\quad\times{\rm Im}\,U^{ph}\left(\frac{q}{2}-k;\rho\right)
\theta \left(\frac{q_0}{2}+k_0\right)
\theta \left(\frac{q_0}{2}-k_0\right)\,,
 \label{phasespace}
\end{eqnarray}
which amounts to neglect the energy and 
momentum dependence of the {\sl p-h} interaction. Finally the imaginary part of
$U^{2p2h}_{L,T}$ will be written as:
\beq
{\rm Im}\, U^{2p2h}_{L,T}(q_0,\vec q;\rho)=
\frac{P(q_0,\vec q;\rho)}{P(m_{\pi},\vec 0;\rho_{\rm eff})}
{\rm Im}\, U^{2p2h}_{L,T}(m_{\pi}, \vec 0; \rho_{\rm eff})
\label{PI2p2h}
\eeq
with $\rho_{\rm eff}=0.75\rho$.

The above formulas refer to homogeneous nuclear matter and provide, through
equations (\ref{Sigma2}) and (\ref{Alpha}), the hypernuclear decay width 
for a $\Lambda$ with momenum $k$ embedded in a constant nuclear density 
$\rho$. The corresponding decay width in finite hypernuclei can be obtained
by applying the local density approximation (LDA).

The latter amounts to consider a {\it local} Fermi momentum 
\beq
k_F(\vec r)=\left\{\frac{3}{2}{\pi}^2\rho 
(\vec r)\right\}^{1/3}\,, 
\label{kFloc}
\eeq
which is defined in the Thomas-Fermi approximation, as follows:
\beq
\epsilon_F(\vec r)+V_N(\vec r)\equiv \frac{k_F^2(\vec r)}{2m_N}+V_N(\vec r)=0.
\label{ThomasFermi}
\eeq
The decay width in finite nuclei is then obtained by:
\beq
{\Gamma}_{\Lambda}=\int d\vec k \,
|\tilde{\psi}_{\Lambda}(\vec k)|^2{\Gamma}_{\Lambda}(\vec k)
\label{Gammloc}
\eeq
where:
\beq
{\Gamma}_{\Lambda}(\vec k)=\int d\vec r \,| {\psi}_{\Lambda}(\vec r)| ^2 
{\Gamma} _{\Lambda}\left[\vec k,\rho (\vec r)\right]
\label{Gammaloc2}
\eeq
and $|\tilde{\psi}_{\Lambda}(\vec k)|^2$ is the ${\Lambda}$ 
momentum distribution.

\section{Theory versus experiment}

In the phenomenological approach for the polarization propagator we have
employed the customary Fermi distribution of the nuclear density:
\beq
\rho_A(r)=\frac{\rho_0}{\left\{\displaystyle 1+
\exp\left[\frac{r-R(A)}{a}\right]\right\}}
\label{nuclden}
\eeq 
with  $R(A)=1.12A^{1/3}-0.86A^{-1/3}$~fm and $a=0.52$~fm.

The ${\Lambda}$ wave function is obtained from a Woods--Saxon 
$\Lambda$--nucleus potential, which exactly reproduces the first two 
single particle eigenvalues ($s$ and $p$ $\Lambda$ levels)
of the hypernucleus under analysis~\cite{ADGR}. 

Short range correlations ($g'$, $g'_{\Lambda}$) are fixed to get agreement 
with experimental data: indeed the Landau parameter $g'$ has been widely
used in the literature, in connection with spin-isospin nuclear responses 
as measured in charge exchange reactions and inelastic electron scattering.
In this context the customary value, used within the RPA framework and 
appropriate to fit the data is $g'=0.6\div 0.7$. 

The available data from $\Lambda$-hypernuclei decay seem to require a 
somewhat larger value of $g'$ (0.8), together with $g'_{\Lambda}\simeq 0.4$.
This is not necessarily in contrast with previous phenomenology, since the
RPA-type correlations are now applied to a richer polarization propagator,
which also includes the 2p-2h excitations. 

\begin{figure}
\begin{center}
\mbox{\epsfig{file=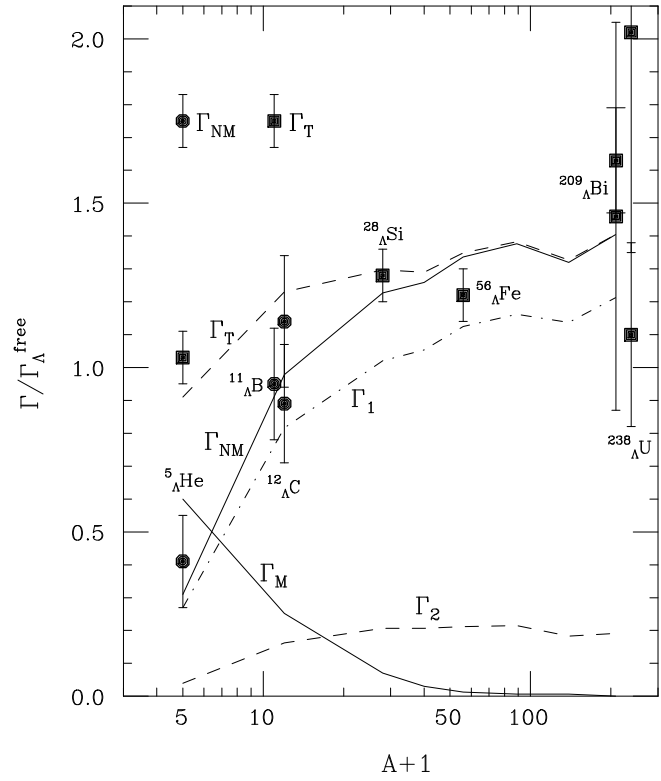,width=.7\textwidth}}
\caption{
Partial ${\Lambda}$ decay widths in finite nuclei as a function of
the nuclear mass number $A$.
The experimental data are taken from Refs.~\cite{Sz91,Ar93,Bh98}}
\label{satu}
\vspace{-1cm}
\end{center}
\end{figure}

In Fig.~\ref{satu} we illustrate the results obtained~\cite{ADGR} with 
the polarization 
propagator method (together with LDA) for the decay widths of various
$\Lambda$-hypernuclei. In addition to the total decay rates, the mesonic 
and non-mesonic partial rates are shown, the latter being separated into the
one-body induced ($\Gamma_1$) and two-body induced ($\Gamma_2$) decay rates.
The total decay rate appears to be rather constant with the mass number,
at least for $A>10$ (saturation property), as a result of a compensation 
between the mesonic decay and the non-mesonic one.

Altogether the results of the theoretical calculation appear to be in good 
agreement with the available experimental data, both on the total rates and,
when available, with the non-mesonic rates. Similar outcomes, though not 
reported here, were obtained with the microscopic calculation of the 2p-2h 
polarization propagator, a fact which demonstrates how the theoretical 
description of these processes is well founded.

\section{The $\Gamma_n/\Gamma_p$ puzzle}

The main problem concerning the weak decay rates 
is to reproduce the experimental value for the ratio ${\Gamma}_n/{\Gamma}_p$
between the neutron-- and the proton--induced widths:
\beq
{\Lambda}n\rightarrow nn\qquad {\mathrm{and}}
\qquad {\Lambda}p\rightarrow np
\label{lambdanp1}
\eeq
Theoretical calculations underestimate the central data for all considered 
hypernuclei:
\beq
\left\{\frac{{\Gamma}_n}{{\Gamma}_p}\right\}^{\rm Th}\ll
\left\{\frac{{\Gamma}_n}{{\Gamma}_p}\right\}^{\rm Exp} ,
\hspace{0.2in}
0.5\lsim \left\{\frac{{\Gamma}_n}{{\Gamma}_p}\right\}^{\rm Exp}\lsim 2
\label{lambdanp2}
\eeq

One should keep in mind that, up to now, the data on the separate rates 
are limited and not precise enough, due to the difficulty in
detecting the products of the non--mesonic decays, especially  neutrons.
The present experimental energy resolution  does not allow to identify the
final state of the residual nuclei in 
$^A_{\Lambda}{\rm Z}\rightarrow {^{A-2}{\rm Z}} + nn$ and
$^A_{\Lambda}{\rm Z}\rightarrow {^{A-2}({\rm Z-1})} + np$. 

In the OPE approximation, by assuming the $\Delta I=1/2$ rule
in the $\Lambda \rightarrow \pi^-p$ and $\Lambda \rightarrow \pi^0n$ 
free couplings, many different calculations 
give small ratios, in the range $0.05\div 0.20$.
\beq
\left[ \frac{\Gamma_n}{\Gamma_p} \right]^{\rm OPE} \simeq 0.05\div 0.20 
\label{lambdanp3}
\eeq
for all the considered systems.

For pure $\Delta I=3/2$ transitions the OPE ratio can increase up to 
about $0.5$. However, this assuption would be inconsistent with the 
fact that the OPE model with $\Delta I=1/2$ couplings well
 reproduces the one--body stimulated non--mesonic rates 
$\Gamma_{1}=\Gamma_n+\Gamma_p$ for light and medium hypernuclei.

Other ingredients beyond the OPE might be
responsible for the large experimental ratios:
\begin{enumerate}
\item 
Calculations with $\Lambda N \rightarrow NN$ transition potentials including
heavy--meson--exchange (e.g. $K$) or direct quark contributions
have improved the situation.\cite{Parreno,Sasaki}
\item
The analysis of the ratio ${\Gamma}_n/{\Gamma}_p$ is influenced by the  
two--nucleon induced process ${\Lambda}NN\rightarrow NNN$: by 
assuming the quasi--deuteron approximation for the absorption of the meson
emitted in the $\Lambda$ decay, the three--body process are mainly 
${\Lambda}np\rightarrow nnp$ and a considerable fraction of  neutrons 
could come from this channel in addition to $\Lambda n \rightarrow nn$ 
and $\Lambda p \rightarrow np$. 
However the inclusion of the new channel would bring to extract from 
the experiment even larger values for the ${\Gamma}_n/{\Gamma}_p$ 
ratios.\cite{ADEM,RaOsSa}
\item
The effect of the final state interaction (FSI) on the spectra of the 
emitted nucleons: the nucleon energy/momentum distributions have been 
calculated~\cite{RamVa} by using a Monte Carlo simulation to describe 
the nucleon rescattering inside the nucleus. The main effects thus obtained
indicate that the nuclear collisions remove nucleons from the high to the 
low energy part of the spectrum, moreover the  
numbers of emitted protons and neutrons tend to become similar (due to charge
exchange processes.
\end{enumerate}
\noindent
A recent upgrading of the calculations quoted above shows that the experimental
proton spectra are compatible with values of the ${\Gamma}_n/{\Gamma}_p$ ratio
between 0.5 and 1.0, still leaving some discrepancy with the direct theoretical
estimates. 

\section{Conclusions}

It is by now well understood that beyond the mesonic channel,
 hypernuclear decay proceeds through non--mesonic processes, induced
by one nucleon or by a pair of correlated nucleons. This channel is  
dominant in medium--heavy hypernuclei, where the Pauli principle strongly
suppresses the mesonic decay.

The mesonic rates have been reproduced quite well by calculations performed 
in different frameworks. 
The non--mesonic rates have been considered within several
phenomenological and microscopic models, most of them based on the pion 
exchange.  
More complex meson exchange potentials and direct quark models have also
been used for the evaluation of non--mesonic decay rates. 
The obtained rates appear to be in agreement with the experimental data.

Although several calculations reproduce the total non--mesonic width,  
$\Gamma_{NM}=\Gamma_n+\Gamma_p (+\Gamma_2)$,
the  obtained $\Gamma_n/\Gamma_p$ are often in strong disagreement
 with the measured central data.
Hence further efforts (especially on the experimental side) must be 
invested in order to understand the detailed dynamics of the non--mesonic 
decay.

Recent experiments at KEK~\cite{KEK} have considerably reduced the error 
bars on  ${\Gamma}_n/{\Gamma}_p$, by means of {\it single} nucleon spectra 
measurements. 
Good statistics coincidence measurements of $nn$ and $np$ emitted pairs 
are required. The angular correlation measurements, as expected from the
forthcoming FINUDA~\cite{Bressani} experiment, will also allow for the 
identification of nucleons  coming out from the  one-- and two--nucleon 
induced processes.


\end{document}